\documentclass[aps,prl,twocolumn,superscriptaddress,showpacs]{revtex4}
\usepackage{graphicx}

\begin{document}

%\preprint{}

\title{Depth dependent spin dynamics of canonical spin glass films: A
low-energy muon spin rotation study}



\author{E.~Morenzoni}
\email[E-Mail:]{Elvezio.Morenzoni@psi.ch} \affiliation{Paul Scherrer
Institut, Labor f\"ur Myon-Spin Spektroskopie, CH-5232 Villigen PSI,
Switzerland}
\author{H.~Luetkens}
\affiliation{Paul Scherrer Institut, Labor f\"ur Myon-Spin
Spektroskopie, CH-5232 Villigen PSI, Switzerland}
\affiliation{Institut f\"ur Physik der Kondensierten Materie, TU
Braunschweig, D-38106 Braunschweig}
\author{T.~Prokscha}
\author{A.~Suter}
\affiliation{Paul Scherrer Institut, Labor f\"ur Myon-Spin
Spektroskopie, CH-5232 Villigen PSI, Switzerland}
\author{S.~Vongtragool}
\affiliation{Paul Scherrer Institut, Labor f\"ur Myon-Spin
Spektroskopie, CH-5232 Villigen PSI, Switzerland}
\affiliation{Kamerlingh Onnes Laboratory, Leiden~University, P.O.B.
9504, 2300 RA  Leiden, The Netherlands}
\author{F.~Galli}
\author{M.B.S.~Hesselberth}
\affiliation{Kamerlingh Onnes Laboratory, Leiden~University, P.O.B.
9504, 2300 RA  Leiden, The Netherlands}
\author{N.~Garifianov}
\affiliation{Kazan Physicotechnical Institute, RAS, Kazan 420029,
Russia}
\author{R.~Khasanov}
\affiliation{Paul Scherrer Institut, Labor f\"ur Myon-Spin
Spektroskopie, CH-5232 Villigen PSI, Switzerland}
\affiliation{Physik Institut der Universit\"at Z\"urich, CH-8057
Z\"urich, Switzerland}

\date{\today}

\begin{abstract}
We have performed depth dependent muon spin rotation/relaxation
studies of the dynamics of single layer films of {\it Au}Fe and {\it
Cu}Mn spin glasses as a function of thickness and of its behavior as
a function of distance from the vacuum interface (5-70 nm). A
significant reduction in the muon spin relaxation rate as a function
of temperature with respect to the bulk material is observed when
the muons are stopped near (5-10~nm) the surface of the sample. A
similar reduction is observed for the whole sample if the thickness
is reduced to e.g. 20~nm and less. This reflects an increased
impurity spin dynamics (incomplete freezing) close to the surface
although the freezing temperature is only modestly affected by the
dimensional reduction.
\end{abstract}


%\keywords{}

\maketitle

%
%------------------- body of paper begins here -------------------------------
%
Spin glasses are founded on the randomness and frustration of the
exchange interaction between diluted spins embedded in a
non-magnetic environment. Their dynamics as a prototype of the
dynamics of glassy and disordered systems with complex phase space
has remained a topical issue for many years
\cite{Binder86,Fisher91,Young98}. Although significant progress has
been made in understanding basic properties of the spin relaxation
above and below the freezing temperature $T_f$, which appears as a
cusp like peak in the zero field cooled (ZFC) susceptibility
\cite{Mydosh93}, the exact nature of the spin glass ground state, of
its ordering, and of the low lying excitations are still
controversial and remain an unsettled theoretical problem
\cite{Palassini00}.

The random freezing of the magnetic moments, which involves strong
cooperative effects, is expected to exhibit finite size effects and
many theoretical investigations have focussed on the question of the
lower critical dimension \cite{Boettcher05}. From the experimental
point of view it is therefore important to investigate whether the
dynamics of the spin glass shows a different pattern if one sample
dimension (e.g. the thickness) is reduced or whether it is
homogenous throughout the sample. The experimental difficulty to
obtain an observable magnetization signal for layer thicknesses in
the nm range has led to many investigations of multilayered samples
consisting of spin glass layers separated by decoupling layers.
Measurements extending down to sub nanometer layers of {\it Cu}Mn
and {\it Ag}Mn found a decrease of $T_f$ below a thickness of 100 nm
with a significant drop only below a few nanometers and still a
finite $T_f$ at one monolayer, indicating that even at this
thickness a canonical spin glass displays the characteristic
macroscopic signatures \cite{Hoines91}. ZFC magnetization and AC
susceptibility measurements have shown a pronounced relative shift
of the frequency dependent $T_f$ upon decreasing the thickness from
bulk to about 2~nm and were interpreted in the frame of crossover
from 3D to 2D dynamics \cite{Sandlund89}.

There are only a few techniques applicable to single layer spin
glasses where possible complications of multilayer samples such as
interlayer diffusion, inhomogeneity and role of decoupling layers
can be avoided. Resistance noise measurements have been used as an
alternative to magnetic measurements \cite{Fenimore99}. Measurements
of the anomalous Hall effect detected a large reduction of the out
of plane magnetization in {\it Au}Fe films (30nm) with respect to
the bulk which was explained as originating from the surface
anisotropy of the impurity spin magnetization introduced by the
presence of the vacuum interface \cite{Seynaeve00}. A new technique
able to provide local information about static fields and dynamic
fluctuations of magnetic moments is now offered by the availability
of a polarized low-energy muon beam at the Paul Scherrer Institute,
which allows depth dependent muon spin rotation/relaxation ($\mu$SR)
investigations in thin single layer spin glasses \cite{Morenzoni04}.

In this letter we report on studies of single layers (10, 20, 50 and
220 nm) of the canonical spin glasses {\it Au}Fe~2.2~and 3 at.\%,
{\it Cu}Mn~2~at.\% and of a double layer {\it Au}Fe(31nm)3~at.\% on
Au(160nm). The polarized muons with their spin precessing and
depolarizing in the neighborhood of the magnetic moments act as
microscopic local probes of statics and dynamics of these moments.
They are very well suited for the study of spin glasses, because of
their high sensitivity in the time window of magnetic fluctuations
of $10^{-4}$-$10^{-10}$~sec~\cite{Uemura85}. The time evolution of
the polarization of the muon ensemble ${P}(t)$ is observed via
detection of the emitted decay positron intensity as a function of
time after implantation. By varying the energy of the low-energy
muons we can follow the temperature dependence of the muon spin
relaxation as a function of layer thickness and of depth below the
surface. We find pronounced thickness and depth dependent effects
that indicate an "incomplete" freezing of the magnetic moments in a
$\sim$ 10 nm region below the surface interfacing the vacuum.

Muon-spin relaxation has provided novel insight about the nature of
magnetic freezing in bulk systems (\cite{Heffner82,Uemura85} and
references therein). The magnetic field at the muon site is mainly
caused by the dipolar fields of the magnetic ions. First experiments
in an external magnetic field transverse (TF) to the initial muon
spin direction showed a rapid relaxation increase when $T_f$ is
approached from above \cite{Murnick76}. At temperatures well above
$T_f$ the electronic moments of the magnetic impurities are rapidly
fluctuating so that the resulting depolarization, given by the field
averaged over the muon life-time (2.2~$\mu$sec), is small. Cooling
the spin glass down to its freezing temperature slows down the
fluctuations, thereby bringing the fluctuation rate into the time
window of the muon and increasing the observable depolarization.
Below $T_f$, the electronic moments become static on the scale of
the muon life-time, causing the muons to precess in the static
dipolar fields of the magnetic ions. In a translational invariant
ferromagnet or antiferromagnet a single precession frequency may be
observed. However, in a spin glass the dipolar fields are random in
magnitude and direction when averaged over the sample as in a
$\mu$SR experiment and only a fast depolarization is observable.
Zero (ZF) and longitudinal field (LF) measurements clearly
demonstrated the coexistence of static and dynamic random local
fields and the gradual build up of a static moment below $T_f$
reflecting homogenous freezing \cite{Uemura85}. Further studies
relating ${P}(t)$ to the local spin autocorrelation function
$\langle\vec{S}(t)\cdot\vec{S}(0)\rangle=q(t)$ and its scaling
behavior as a function of the applied field showed that above and
close to $T_f$ the correlations of the impurity moments are strongly
non-exponential \cite{Keren00}.

Our samples (size 25 x 25 mm$^{2}$) have been prepared by
co-sputtering high purity metals onto a Si/SiO$_{2}$ substrate.
Before the argon sputter gas was admitted the equipment was pumped
down to UHV conditions ($10^{-8}$~mbar) to avoid oxidation of the
films during fabrication. The films were analyzed using Rutherford
backscattering (RBS) and electron microprobe analysis. RBS confirmed
a homogenous depth distribution of the magnetic impurities.
Simultaneously, narrow stripes for resistivity and Hall-effect
measurements for $T_f$ determination were sputtered. Thicker films
prepared for magnetic susceptibility measurements confirmed the
freezing temperatures reported in the literature for the
concentration chosen. Within the error the resistivity of the films
(about twice as large as in the bulk material) does not depend on
the thickness indicating the equivalence of the disorder in all
films. Muons with energies between 1 and 22.5~keV were implanted in
the samples. The range profiles were calculated using the Monte
Carlo program TRIM.SP \cite{Eckstein94,Morenzoni02}. Over the
investigated temperature range the muons do not diffuse and randomly
occupy octahedral interstitial sites in the fcc-lattice. We
performed the experiments in ZF and in 10 mT (TF). A transverse
field influences properties of the spin glass such as the sharpness
of the transition, but leaves unaltered the essential features
reflecting the depth and thickness dependent dynamical behavior of
the films. Also previous $\mu SR$ and NMR investigations are
consistent with an essentially field independent spectral function
$J(\omega)$ of the fluctuating fields \cite{MacLaughlin83}. The
observed ${P}(t)$ was analyzed by fitting with a stretched
exponential relaxation function, $\exp{[-(\lambda t)^\beta}]$
(multiplied by a precession factor $\cos {(\omega_\mu t)}$ for the
TF measurements). This function has been found appropriate in cases
where a complex non-exponential fluctuation pattern is expected
\cite{Campbell94}. Additionally we analyzed the ZF measurements with
a muon spin relaxation function, which is able to describe the
coexistence below $T_f$ of static and dynamic random fields produced
by the impurity moments \cite{Uemura85}:

\begin{eqnarray}
P(t)=P(0)\left [ \frac{1}{3} e^{-\sqrt{\lambda_d t}}
          + \frac{2}{3}
        \left (1-\frac{a_s^2t^2}{\sqrt{\lambda_d t+a_s^2t^2}} \right )
        e^{-\sqrt{\lambda_d t+a_s^2t^2}}\right ] \label{spinglass}.
\end{eqnarray}

\begin{figure}[h]
   \centering
   \includegraphics[width=0.85\linewidth]{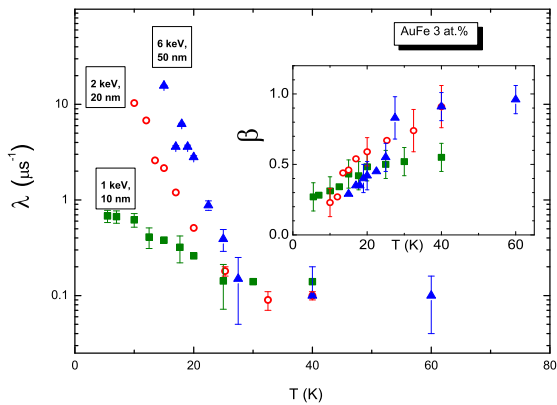}
   \vspace{-0.0cm}
   \caption{Muon depolarization rate ($\lambda$) and stretch exponent
   ($\beta$, see inset) versus temperature for {\it
Au}Fe thin films of 10, 20 and 50~nm, 10 mT TF data. Muons of 1, 2,
6 keV respectively, are stopped in the center of the layer.}
   \label{AuFe-1}
\end{figure}

Here $a_s=a_0 \sqrt{Q}$ is the static field width probed by the
muons, $a_0$ is the width corresponding to full static moments at
T=0 and $Q$ is the Edwards-Anderson order parameter describing the
non vanishing part of the impurity spin autocorrelation function for
infinite time \cite{Fisher91}. The dynamical relaxation introduced
by the fluctuations is described by the parameter $\lambda_d$. The
results for the muon spin relaxation rate $\lambda$ and exponent
$\beta$ as a function of temperature are shown in Fig. \ref{AuFe-1}
for single layers of {\it Au}Fe  of various thicknesses (the {\it
Cu}Mn system gives similar results). The effect of the reduced
sample dimension is evident. \vspace{-0cm}
For each thickness, $\lambda(T)$ increases steadily on approaching
$T_f$ from above, reflecting the slowing down of the moments. In the
50 nm sample the increase is similar to the one observed for bulk
samples. However, already in the 20 nm layer a reduction of
$\lambda(T)$ is visible. This reduction is much more pronounced in
the 10 nm layer so that well below $T_f$ $\lambda \lesssim 1 \mu
s^{-1}$ less than 10\% of the 50 nm value; this factor cannot be
accounted for by a pure geometric effect on the static width of the
spin glass surface. The exponent $\beta$ reaches 1 (corresponding to
a single correlation time) only well above $T_f$ and decreases
towards $\frac{1}{3}$ at low temperatures, a behavior qualitatively
observed in bulk samples. The large reduction of the relaxation rate
in the thin films points to the persistence of spin fluctuations at
low temperatures. This can be due to a finite size or to a surface
effect. In the latter case the effect should emerge also in a
thicker sample. We have therefore performed depth dependent
investigations on {\it Au}Fe~3~at.\% 50 and 220 nm films by varying
the implantation energy between 1 keV (mean depth \={d}=6 nm) and
22.5 keV (\={d}=68 nm).
\vspace{-0cm}
\begin{figure}[htb]
  \centering
  \includegraphics[width=0.85\linewidth]{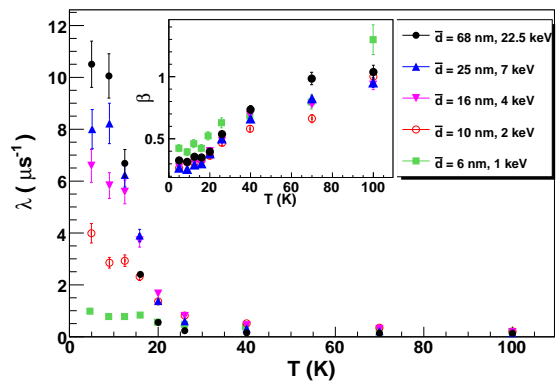}
  \vspace{-0cm}
  \caption{Muon depolarization rate ($\lambda$) and stretch exponent
   ($\beta$, see inset) as a function of temperature (T) at different (mean) depths
   in {\it Au}Fe~3~at.\%. Thickness: 220 nm, ZF data.}
  \label{AuFe_keV}
\end{figure}

Results are plotted in Fig. \ref{AuFe_keV}. Clearly, $\lambda$ is
depth dependent. A strong reduction is found at 2 keV and below
corresponding to stopping profiles centered in the top 10 nm of the
spin glass. Again similar behavior is observed in {\it Cu}Mn. The
results imply that the reduction of the muon spin relaxation in the
20 nm and the strong suppression in the 10 nm sample are a
manifestation of the dominant contribution of the surface layer at
these thicknesses. In principle a smaller $\lambda$ in the TF and ZF
measurements can be due to a pure reduction of the static width
below $T_f$ or to enhanced dynamics and consequent motional
narrowing or a combination of both. At the moment our apparatus does
not allow routine LF investigations in a large range of fields to
follow the repolarizing effect typical of the pure static case.
However, LF measurements in the 10 nm {\it Cu}Mn sample in fields up
to $\sim$ 13 mT at 4 K do not show any change of $P(t)$. The fact
that this LF does not decrease relaxation rates of $\lesssim 1 \mu s
^{-1}$ (corresponding to $\lesssim$ 1 mT static field width)
indicates dynamical effects as a cause of the observed behavior.
This is also confirmed by an heuristic interpretation of the $\beta$
and $\lambda$ parameters. Formally, a stretched exponential $P(t)$
can be expressed as a distribution of exponential muon relaxation
rates $\lambda_i$ each reflecting the contribution of local spins
relaxing with a time $\tau_i \propto \lambda_i$ ~\cite{Campbell94}.
The distribution function of $\lambda_i$ reflects the weighting
distribution $g(\tau_i)$ of the corresponding exponential
autocorrelation functions with $\beta$ and $\lambda$ determining the
shape of $g(\tau_i)$. Reducing $\beta$ below 1 widens the
distribution. In addition lowering $\lambda$ shifts the weight from
long correlation times, which appear as static component to the
$\mu^+$, to very short ones. The depth dependence of $\lambda$ for
$T< T_f$ reflects therefore a wide distribution of spin relaxation
times with persistence of a substantial dynamic component close to
the surface.
%
%\vspace{-0.0cm}
\begin{figure}[hb]
 \includegraphics[width=0.85\linewidth]{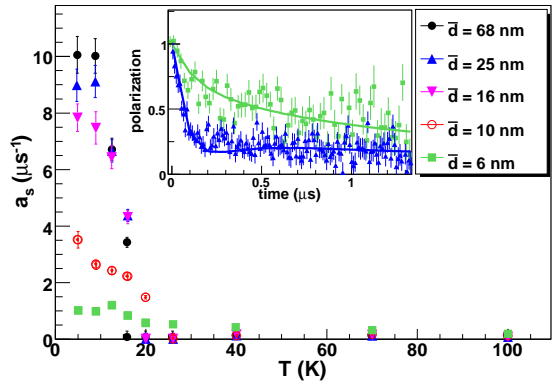}%
 \vspace{-0.0cm}
 \caption{\label{AuFe_a_s}
 Static amplitude of random fields $a_s$ in the {\it Au}Fe film
 (thickness 220 nm) at different depths . In the inset polarization spectra (normalized to the asymmetry
 obtained in pure Au) at 5K and 2 different depths are shown. The curves
 are fit to Eq. \ref{spinglass}.
}
\end{figure}

The analysis of the ZF data with Eq. \ref{spinglass} allows to
separate static from dynamic contributions. A pronounced depth
dependence is observed in the magnitude of the static field
component, $a_s$ (Fig. \ref{AuFe_a_s}), which attains finite values
below $\sim T_f$ to saturate on lowering the temperature. Overall
the curve shows that this order parameter is gradually reduced in
the $\sim$ 10 nm layer below the surface. This is clearly visible in
the shape of $P(t)$ taken at 5 K and two different depths of the 220
nm film (Fig. \ref{AuFe_a_s} inset). Whereas well below the surface
$P(t)$ exhibits a rapid decrease followed by a tail characteristic
of the presence of (pronounced) static field components, close to
the surface weak relaxation dominates the spectrum. The dynamic
component $\lambda_d$ peaks around $\sim T_f$ in agreement with bulk
results \cite{Uemura85}. The reduction of $a_s$ below $T_f$ and the
behavior of $\lambda_d$ means that rapid large fluctuations with
typical correlations times $\tau << \frac{1}{a_s}=10^{-6} s$ around
a preferred direction do not freeze even at the lowest temperature.
It should be remarked that the increased dynamics is not accompanied
by a concomitant sizeable reduction of $T_f$. This is visible in
Fig. \ref{AuFe_a_s} which evidences the strong variation of $a_s$
with depth and a within errors practically unchanged $T_f$ as
determined from $a_s(T_f)=0$. Also in spin glasses of {\it Au}Fe,
{\it Ag}Mn and {\it Cu}Mn at higher magnetic concentration the
reduction of the thickness  to 10~nm was found to reduce the
freezing temperature by at most 30\% \cite{Fenimore99,Hoines91}.
Seynaeve et al. \cite{Seynaeve00} observed a strong decrease of the
magnitude of the anomalous Hall effect as the thickness of the spin
glass gets smaller and a disappearance of the effect when their {\it
Au}Fe film was sandwiched between layers of pure Au. Both
observations were interpreted as the result of in-plane surface
anisotropy of the static Fe-moments caused by spin-orbit scattering
of electrons by the non-magnetic host atoms. Our $\mu SR$
measurements of a double layer {\it Au}Fe(31~nm)/Au(160~nm) also
exhibit a reduction of the low-temperature relaxation near the {\it
Au}Fe~/~vacuum interface, but not near the {\it Au}Fe/~Au interface.
However, surface anisotropy as the only cause of the observed
thickness and depth behavior can be ruled out since in the dilute
case randomly oriented or aligned dipoles both produce Lorentzian
field distributions with the same width \cite{Berkov96} and would
leave the $\mu SR$ spectra and parameters unmodified. We confirmed
this by numerical simulations. On the other hand our results,
showing the presence of a fluctuation spectrum with large amplitude
oscillation below $T_f$, can explain the Hall effect measurements of
\cite{Seynaeve00}.

In conclusion we have used the sensitivity and the depth profiling
capacity of LE-$\mu SR$ to study the inhomogeneous magnetization and
dynamic profile in dilute magnetic alloys. We find that on a length
scale of $\sim$ 10 nm the surface of canonical spin glass systems
exhibits a considerably enhanced fluctuating behavior in comparison
with the bulk, whereas the apparent freezing temperature does not
change dramatically. This reduction of the order parameter modifies
the magnetic state of films of comparable thickness. To our
knowledge no theoretical model is able to predict this inhomogeneous
behavior or to identify the underlaying physical mechanism. However,
recent calculations of the low energy excitations of the
Edwards-Anderson model of Ising spin glasses find that a dimensional
reduction of 10-30\% of $T_f$ is accompanied by a disproportionately
larger reduction of the stiffness exponent until the lower critical
dimension d=$\frac{5}{2}$ is reached \cite{Boettcher05}. Since this
exponent governs the typical energy scale for excitation, lowering
it reduces the resistance against the formation of low energy
excitations, a behavior in qualitative accordance with our
observation.

It has been predicted that in an inhomogeneous situation such as the
one represented by the presence of a vacuum interface the RKKY
interaction, which determines the interaction between the impurity
spins, may be modified \cite{Helman94} but its effect on the spin
dynamics or freezing behavior remains to be quantified. Recent
experiments \cite{Sirotti00} have shown that the surface of
ferromagnetic layers is weakly exchange coupled to the bulk,
resulting in faster dynamics of the surface magnetization but
contrary to the spin glass in the ferromagnetic case only a surface
layer of sub nm thickness is concerned. On the other hand, the
length scale of enhanced fluctuations observed here may explain why
in previous multilayer experiments decoupling layers of several tens
of nm were necessary to observe dimensional effects
\cite{Granberg91}. We hope that the results presented here will
stimulate theoretical calculations of the spin dynamics in the
situation where translational invariance is not satisfied.

This work was performed at the Swiss Muon Source S$\mu$S, Paul
Scherrer Institute, Villigen, Switzerland and financially supported
by FOM, The Netherlands. We thank G. Nieuwenhuys for the important
contributions in the early stage of this work.
\bibliographystyle{prsty}
%\bibliography{tp_bib}

\end{document}